\documentclass[11pt]{article}

\usepackage{arxiv}

\usepackage[utf8]{inputenc} % allow utf-8 input
\usepackage[T1]{fontenc}    % use 8-bit T1 fonts
\usepackage[colorlinks,urlcolor=blue]{hyperref}       % hyperlinks
\usepackage{url}            % simple URL typesetting
\usepackage{booktabs}       % professional-quality tables
\usepackage{amsfonts}       % blackboard math symbols
\usepackage{nicefrac}       % compact symbols for 1/2, etc.
\usepackage{microtype}      % microtypography
\usepackage{lipsum}		% Can be removed after putting your text content
\usepackage{graphicx}
\usepackage{doi}

\usepackage{wrapfig}

\hypersetup{pdfborder = 0 0 0}

\usepackage{notoccite}
\usepackage[super,numbers,sort&compress]{natbib}

\usepackage{hyperref}
\usepackage{xcolor}
\hypersetup{
    colorlinks = true,
    linkcolor = blue,
    citecolor = blue,
    urlcolor = blue
}

\title{A mechanistic model for smallpox transmission via inhaled aerosols inside respiratory pathways}

\author{ \href{https://orcid.org/0000-0003-1464-8425}{\includegraphics[scale=0.06]{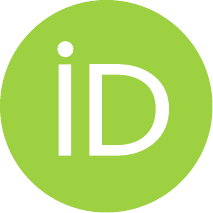}\hspace{1mm}Saikat~Basu}\thanks{Corresponding author. Web: \href{https://www.sdstate.edu/directory/saikat-basu}{https://www.sdstate.edu/directory/saikat-basu}} \\
	Department of Mechanical Engineering\\
	South Dakota State University\\
	Brookings, SD 57007, United States\\
	\texttt{Saikat.Basu@sdstate.edu} \\
	\And
	Abir Malakar \\
	Department of Mechanical Engineering\\
	South Dakota State University\\
	Brookings, SD 57007, United States\\
	\texttt{Abir.Malakar@sdstate.edu} \\
 \And
	Mohammad Mehedi Hasan Akash \\
	Department of Mechanical Engineering\\
	South Dakota State University\\
	Brookings, SD 57007, United States\\
	\texttt{Mohammad.Akash@sdstate.edu} \\
}

\renewcommand{\shorttitle}{On modeling airborne smallpox transmission}

\hypersetup{
pdftitle={Mechanics modeling for airborne smallpox transmission},
pdfsubject={fluid dynamics},
pdfauthor={Saikat Basu, Abir Malakar, Mohammad Mehedi Hasan Akash},
pdfkeywords={Airborne transmission, Fluid dynamics modeling, Respiratory transport, Smallpox},
}

\begin{document}
\maketitle

\begin{abstract}
	Investigations on airborne transmission of pathogens constitute a rapidly expanding field, primarily focused on understanding the expulsion patterns of respiratory particulates from infected hosts and their dispersion in confined spaces. Largely overlooked has been the crucial role of fluid dynamics in guiding inhaled virus-laden particulates within the respiratory cavity, thereby directing the pathogens to the infection-prone upper airway sites.  Here, we discuss a multi-scale approach for modeling the onset parameters of airway infection based on flow physics. The findings are backed by Large Eddy Simulations of inhaled airflow and computed trajectories of pathogen-bearing aerosols/droplets within two clinically healthy and anatomically realistic airway geometries reconstructed from computed tomography imaging. As a representative anisotropic pathogen that can transmit aerially, we have picked smallpox from the Poxviridae family to demonstrate the approach. The fluid dynamics findings on inhaled transmission trends are integrated with virological and epidemiological parameters for smallpox (e.g., viral concentration in host ejecta, physical properties of virions, and typical exposure durations) to establish the corresponding infectious dose (i.e., the number of virions potent enough to launch infection in an exposed subject) to be, at maximum, of the order of $\mathcal{O}(2)$, or more precisely 1 to 180. The projection agrees remarkably well with the known virological parameters for smallpox.
\end{abstract}

% keywords can be removed
\keywords{Airborne transmission, Fluid dynamics modeling, Respiratory transport, Smallpox, Virology}

\section{Introduction}
%\setlength{\parindent}{10pt}
%\vspace{-3mm}
The physical parameters governing inter-individual aerial transmission of pathogens are deeply rooted in the mechanics of the ambient airflow fields and that of the transmitting droplets and aerosols\cite{stilianakis2010jrs}. The problem is nuanced and is being studied across different length and time scales. 
Once aerosols (typically < $5$ $\mu$m) and droplets ($\ge$ $5$ $\mu$m)\cite{who2014} laden with pathogens manage to enter a subject's respiratory airway, transport is significantly impacted by the complex morphology of the anatomical flow path, the interplay between the inertial motion of the particulates and the surrounding airflow, and finally by the rheology of the mucociliary system\cite{basu2022ro}. A detailed fluid mechanics-governed understanding of inhaled particle transport within the upper respiratory tract (URT) and the resulting transmission trends of pathogens embedded in such particulates can help quantify key basic parameters related to the process of infection onset, e.g., the pathogen-specific \textit{infectious dose} ($I_D$), representing the minimum number of pathogens that can initiate  infection\cite{zwart2009prs}.

\begin{figure}[t]
\begin{center}
\includegraphics[width=7.5cm]{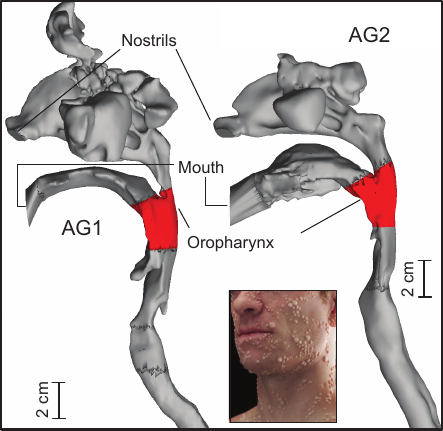}
\caption{Oropharynx, colored red in the two test geometries (labeled as anatomic geometry 1 or AG1 and anatomic geometry 2 or AG2), is the initial infection trigger site for smallpox in the upper respiratory tract. Inset shows an external illustration of smallpox outbreak, adapted from Getty Images\textsuperscript{\textregistered}. The 2-cm scale bars on the left and right are respectively for AG1 and AG2.}\label{fig1}
\end{center}
\end{figure}

The modeling for airborne infection onset demonstrated in this study builds on computational fluid dynamics simulations of inhaled airflow and particle transport inside medical scan-based anatomically realistic URTs. Herein, the particles bear the physical properties of pathogen-embedded aerosols and droplets formed from a host's respiratory ejecta. As a specific example, we track down the $I_D$ for smallpox virus (\textit{variola} virus) of the \textit{Poxviridae} family, by identifying the inhaled particulate sizes 
that land directly at the dominant infective site (oropharynx\cite{milton2012frontier}, see Fig.~\ref{fig1}) along the URT, thus ferrying the virions there.
Note that smallpox is an anisotropic virus that can spread via aerial inhalation\cite{milton2012frontier} and although it was declared eradicated by the WHO in 1980, our findings can help establish a foundational mechanics-guided paradigm for \textit{in silico} exploration of related emerging pathogens, such as the monkeypox\cite{parker2007fm}. The generic modeling approach for tracking airway infection onset presented here is comparable to our published study\cite{basu2021scirep,basu2021aps} on SARS-CoV-2, which accurately quantified the relevant $I_D$ to be $\approx$ 300, with subsequent validation from multiple \textit{in vivo} investigations\cite{ryan2021nature,karimzadeh2021ei,prentiss2022plos} with human and animal models.

\section{Methods}
%\setlength{\parindent}{10pt}
%\vspace{-3mm}

\subsection{Digital fabrication of anatomically realistic upper airway tracts}
We have reconstructed the airspace domain from high-resolution medical-grade computed tomography (CT) scans derived from healthy subjects to generate two normal URTs (labeled as Anatomic Geometry 1 or AG1 and Anatomic Geometry 2 or AG2; see Fig.~\ref{fig1}). The CT images offered a resolution of < 0.4 mm and the airspace segmentation required a radiodensity threshold of -1024 to -300 Hounsfield units. To confirm anatomical authenticity of the test geometries, we also obtained post-reconstruction feedback from practicing rhinologists. For simulations, the spatial domains were segregated into > 6 million unstructured, graded, tetrahedral elements with 4 layers of prism cells of combined 0.1 mm thickness along the cavity walls. 

The spatial meshing parameters are backed by earlier studies on the effects of mesh refinement on computational simulation outcomes in comparable URT systems\cite{basu2017mesh,frank2016jampdd}. For additional details on the recreation of anatomic realism associated with healthy upper airway domains (for subjects exposed to an airborne pathogenic transmission and subsequent infection onset), see our earlier works\cite{perkins2018ohns,basu2018ijnmbe,kimbell2019lsm,kimbell2018rdd,farzal2019ifar,tracy2019ifar}. For representative experimental validation of the computational modeling approach described next, the reader is directed to our recent publications\cite{basu2020scirep,akash2023fdd} involving artificial spray experiments conducted in 3D-printed anatomical casts developed from CT-based airway reconstructions.

\subsection{Simulation of inhaled transport} 
\begin{figure}[t]
\begin{center}
\includegraphics[width=7.5cm]{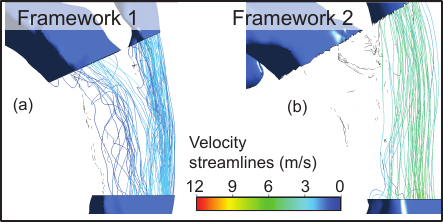}
\caption{Representative simulated velocity streamlines extracted from the inhaled airflow near the oropharynx, respectively for Framework 1 (in panel a) and Framework 2 (in panel b). }\label{fig2}
\end{center}
\end{figure}

To replicate normal breathing parameters\cite{garcia2009dosimetry,he2013ast}, we have simulated inhalation rates of $15$ and $30$ L/min under pressure-gradient driven transient flow conditions, using Large Eddy Simulation (LES) scheme. Subgrid scale dynamics was resolved using the Kinetic Energy Transport Model\cite{baghernezhad2010jt} and the solutions followed second order approximation with convergence determined by minimizing the velocity and mass continuity residuals. For the simulated fluid medium being passed into the test URTs, we used physical properties of inhaled warmed-up air, namely $1.204$ kg/m$^3$ for density and $1.825 \times 10^{-5}$ kg/m.s for viscosity coefficient. The airflow simulation duration covered $0.35$ s, with time-steps of $0.0002$ s. Note that we employed two contrasting frameworks on the air inlet conditions: (a) Framework 1, with air being inhaled through both nostrils and mouth, in AG1; and (b) Framework 2, with air being inhaled only through the nostrils, in AG2. See Fig.~\ref{fig2}a-b for the respective modeling frameworks.
Against the post-convergence ambient airflow, the transport of inhaled aerosols and droplets was mimicked by tracking inert particles with a Lagrangian discrete phase model that accounted for effects such as drag and Saffman lift force. The material density of the particles was $1.3$ g/ml, similar to that of post-emission environmentally dehydrated respiratory ejecta\cite{stadnytskyi2020pnas} from a host that is now being inhaled by an exposed subject. The simulated particle sizes were monodisperse, with diameters $\in [0.1, 55]$~$\mu$m, and were let into the numerical domain through nostrils. Numbers of tracked particles were 1660 in AG1 and 2370 in AG2 for each diameter and were correlated to the nostril cross-sections of the subjects. The maximum size threshold ensured that we do not consider droplets that would anyway undergo prompt gravitational sedimentation in the outside air\cite{stadnytskyi2020pnas,bourouiba2021arfm,bourouiba2021arbe,bourouiba2020jama}.

\subsection{Connecting the fluid dynamics findings to virological and epidemiological parameters}\label{integration} 
Oropharynx, marked in Fig.~\ref{fig1}, is the dominant initial infection site for smallpox virus\cite{milton2012frontier}. From post-processing the computationally simulated inhaled transmission trends, we can figure out the deposition efficiency for each particle size at the oropharynx; e.g., see Fig.~\ref{fig3}a-b. Let the efficiency be $\mathcal{E}_i$ (in \%) for particle diameter $i$. From published studies\cite{xie2009jrsi}, we have extracted the size distribution of respiratory ejecta particulates that get inhaled by an exposed subject; for a similar adaptation of particle size partitions in our recent published work, see\cite{basu2021scirep}. Assume $n_i$, modulated by the size distribution function, to be the number of particles of size $i$ inhaled in each unit time. Further, the exposure duration that has led to confirmed infection onset has been reported in literature, let it be represented as $T \in (1.7, 16.7)$ hours\cite{milton2012frontier}. So, during the prescribed exposure time, the number of inhaled particulates of size $i$ landing at the oropharynx could be mathematically estimated as:~
\begin{equation}\label{e1}
\mathcal{N}_i = \mathcal{E}_i n_i T.
\end{equation}

To quantify the number of virions ferried by the deposited particulates at the oropharynx, consider now the weight of each virion $w$; it has been reported as $5$ -- $10$ fg\cite{johnson2006sabc}. Furthermore, the viral load concentration in oral samples of smallpox hosts has been shown to be $10^2$ -- $10^6$ fg/$\mu$l\cite{sofi2003jcm}; let it be $\zeta$ in consistent units. This implies that the number of virions embedded in unit volume of respiratory ejecta is $\zeta/w$. Therefore, the number of virions transmitted to the oropharynx via inhalation during the infection-launching exposure time is:
\begin{equation}\label{e2}
I_D = \sum_{i}^{}\frac{i^3 \pi \mathcal{N}_i \zeta}{6w}
\end{equation}

\section{Results}
\begin{figure}[t]
\begin{center}
\includegraphics[width=\textwidth]{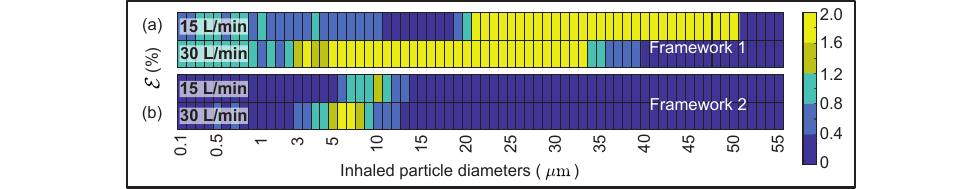}
\caption{(a, b) Heatmaps for inhaled oropharyngeal deposition efficiency $\mathcal{E}$, calculated as the percentage of the tracked aerosols/droplets (that entered through the geometry inlets) depositing at the oropharynx. The inhaled particulate sizes are on the horizontal axis. The different rows, as marked, are for the discrete inhaled airflow rates tested, i.e., 15 and 30 L/min. Therein, top two rows correspond to simulated particle deposition efficiency at the oropharynx using Framework 1. The bottom two rows correspond to simulated particle deposition efficiency at the oropharynx using Framework 2.}\label{fig3}
\end{center}
\end{figure}

%\setlength{\parindent}{10pt}
%\vspace{-3mm}
In Framework 1 (where air is inhaled through both mouth and nostrils), the oropharyngeal space sees a chaotic mixing of the oral and nasal fluxes (as exemplified in Fig.~\ref{fig2}a), resulting in lower air speeds in that region, compared to those in Framework 2 (where air is only coming in through the nostrils). The flux comparison, with 50 randomly selected velocity streamlines, is shown in Fig.~\ref{fig2}a-b. The higher inertia of the flow field in Framework 2 drives the inhaled particles further downstream, resulting in less oropharyngeal deposition (compared to that in Framework 1) per unit time. Figure~\ref{fig3}a-b demonstrate the corresponding heatmaps for the deposition efficiency of each inhaled particle size at the oropharyngeal walls. Clearly, $\mathcal{E}_i$ in Framework 2 is lower.

\subsection{In silico estimation of smallpox infectious dose} 
By invoking equations~\ref{e1} and \ref{e2}, Framework 1 results in an $I_D$ range of $1$ -- $\mathcal{O}(5)$, while Framework 2 leads to an $I_D$ of $1$ -- $\mathcal{O}(2)$, or more precisely min$\{I_D\} \in (1, 180]$. The estimation considers the overall span of virological measurements (described in \S\ref{integration}) as inputs to the particle transmission data derived from the simulations.

\section{Conclusion}
%\setlength{\parindent}{10pt}
%\vspace{-3mm}
The $I_D$ projected from Framework 2, with open nostrils allowing inhaled air flux and with a closed mouth, suggests that a range of $1$ to $180$ virions could be sufficiently potent to trigger a smallpox infection through aerial inhalation. This aligns remarkably well with established\cite{arizona2024} \textit{in vivo} smallpox $I_D$ estimates of $10$ to $100$. Consequently, a crucial insight is that, as an \textit{in silico} computational modeling approach, Framework 1 (i.e., allowing air intake through both nose and mouth) yields inaccurate $I_D$ assessments. In retrospect, this discrepancy is justified as mouth inhalation constitutes < $10$\% of breathing time in healthy adults\cite{camner1980er}. Hence as a final note, the modeling platform exemplified by Framework 2 effectively integrates fluid dynamics findings with the virological parameters for a reliable estimate of smallpox infectious dose and the translational technique can be extended\cite{lu2022boh} to the trasmission modeling of other pathogens of the pox family, such as the monkeypox\cite{aljabali2022mp,parker2007fm}.\\

\noindent\textbf{Acknowledgements:} Preliminary results from this work were presented at the American Physical Society's Annual Meetings of the Division of Fluid Dynamics\cite{abir2023aps,akash2022apssp} and are in review for presentation at the International Congress of Theoretical and Applied Mechanics (to be held at Daegu, South Korea in 08/2024). The authors also acknowledge Dr.~Azadeh Borojeni (Postdoctoral Fellow at the \href{https://www.sdstate.edu/directory/saikat-basu}{Basu Lab}, South Dakota State University), Mr.~Aditya Tummala (High School Intern at the \href{https://www.sdstate.edu/directory/saikat-basu}{Basu Lab}, South Dakota State University), and Dr.~Arijit Chakravarty (Chief Executive Officer, Fractal Therapeutics, Lexington, MA) for initial conceptual discussions on the topic.

\noindent\textbf{Funding:} This material is based upon work supported by the National Science Foundation CAREER Award under \href{https://www.nsf.gov/awardsearch/showAward?AWD_ID=2339001&HistoricalAwards=false}{Grant No.~2339001}, with SB as the Principal Investigator.\\

\bibliography{NSF_PIPP_v1}

%\begin{thebibliography}{1}

%\bibitem{i}
%Milton D.K.: What was the primary mode of smallpox transmission? Implications for biodefense. \textit{Frontiers in Cellular and Infection Microbiology}, \textbf{2}:1--7, 2012.

%\bibitem{ii}
%Basu S.: Computational characterization of inhaled droplet transport to the nasopharynx. \textit{Scientific Reports} \textbf{11}:1--13, 2021.

%\bibitem{iii}
%Xie X. et al.: Exhaled droplets due to talking and coughing. \textit{Journal of the Royal Society Interface}. \textbf{6}:S703-S714, 2009.

%\bibitem{iv}
%Johnson L. et al.: Characterization of vaccinia virus particles using microscale silicon cantilever resonators and atomic force microscopy. \textit{Sensors and Actuators B: Chemical}, \textbf{115(1)}:189--197, 2006.

%\bibitem{v}
%Sofi I.M. et al.: Real-time PCR assay to detect smallpox virus. \textit{Journal of Clinical Microbiology}. \textbf{41(8)}:3835--3839, 2003.

%\bibitem{vi}
%Arizona Department of Health Services 2004, web site  \textsf{\scriptsize https://www.azdhs.gov/documents/preparedness/emergency-preparedness/zebra-manual/zm-s4-smallpox.pdf}

%\end{thebibliography}

\end{document}